\documentclass{ifacconf}

\usepackage{graphicx}      
\usepackage{natbib}        
\usepackage{amsmath,color} 
\usepackage{amssymb}  
\usepackage{makecell}
\usepackage{booktabs}
\usepackage{soul,comment}
\usepackage{xcolor}
\usepackage{multirow}
\usepackage{url}

\allowdisplaybreaks

\begin{document}
\begin{frontmatter}

\title{\large Modeling and Control of Diesel Engine Emissions using Multi-layer Neural Networks and Economic Model Predictive Control} 


\author[First]{Jiadi Zhang}, 
\author[First]{Xiao Li}, 
\author[Second]{Mohammad Reza Amini}, 
\author[First]{Ilya Kolmanovsky},
\author[Third]{Munechika Tsutsumi}, 
\author[Third]{Hayato Nakada}

\address[First]{Department of Aerospace Engineering, University of Michigan, Ann Arbor, MI 48109, USA (e-mails: \{jiadi,hsiaoli,ilya\}@umich.edu)}
\address[Second]{Department of Naval Architecture and Marine Engineering, University of Michigan, Ann Arbor, MI 48109, USA (e-mail: mamini@umich.edu)}
\address[Third]{Hino Motors, Ltd., Tokyo 191-8660, Japan ({e-mails: \{mu.tsutsumi,hayato.nakada\}@hino.co.jp})}

\begin{abstract}    
This paper  
presents the results of developing a multi-layer Neural Network (NN) to represent diesel engine emissions and integrating this NN into control design.
Firstly, a NN is trained and validated to simultaneously predict oxides of nitrogen ($NOx$) and $Soot$ using both transient and steady-state~
data. 
Based on the input-output correlation analysis, inputs to NN with the highest influence on the emissions are selected while keeping the NN structure simple. Secondly, a co-simulation framework is implemented to integrate the NN emissions model with a model of a diesel engine airpath system built in GT-Power and used to 
identify 
a low-order linear parameter-varying (LPV) model for emissions prediction. Finally, an economic supervisory model predictive controller (MPC) is developed using the LPV emissions model to adjust setpoints to an inner-loop airpath tracking MPC.~
Simulation results are reported illustrating the capability of the resulting controller 
to  reduce $NOx$,  meet the target $Soot$ limit, and track the adjusted intake manifold pressure and exhaust gas recirculation (EGR) rate targets.
\end{abstract}


\end{frontmatter}

\section{Introduction}\vspace{-4pt}
To 
achieve emission reduction in diesel engines, advanced engine control methods can be leveraged to coordinate engine fueling, exhaust gas recirculation (EGR) valve, and variable geometry turbocharger (VGT). 
In particular, Model Predictive Control (MPC) has shown significant promise for improving engine transient control while satisfying state and control constraints (\cite{norouzi2021model}). Accurate and low-complexity engine-out emission models 
can support control design and verification before physical engine testing 
(\cite{li2017emissions}).

Different emission modeling methods have been proposed in the literature including map-based quasi-steady models (\cite{hagena2006transient}), physics-based grey-box models (\cite{heywood2018internal,arregle2008sensitivity}), and machine learning based models (\cite{winkler2010comparison,prokhorov2008neural,de2011neural,li2017emissions}). 
As shown in \cite{hagena2006transient}, map-based models may not accurately predict transient effects such as smoke spikes during fuel tip-ins. On the other hand, physics-based grey-box models often exploit in-cylinder pressure as an input, which is not available in commercial vehicles in production. Hence, in this paper, a data-driven  machine learning-based model is developed using limited training data to support the emissions-oriented MPC design.

The existing literature on the application of MPC to diesel engines 
has addressed intake manifold pressure and airflow/EGR rate setpoint tracking in the airpath system (\cite{huang2013rate,huang2018toward,ortner2007predictive}) as well as higher-level objectives such as enhanced fuel economy or reduced emissions (\cite{huang2020energy}).
\cite{liao2020model,liu2021simultaneous} have developed solutions along the lines of  nonlinear economic MPC (EMPC) to control $NOx$ and $Soot$ emissions. \cite{broomhead2016economic} have used EMPC for power tracking while limiting emissions in diesel generator applications.

In this paper we focus on  a hierarchical MPC architecture for emissions reduction with an outer-loop (supervisory) economic MPC which adjusts fueling rate and intake manifold pressure and EGR rate targets 
(set-points)  to an inner-loop tracking MPC for the airpath system.  For the inner-loop MPC, we adopt the airpath controller design from \cite{liao2020model,ZHANG2022181},
which exploits feedforward MPC and feedback rate-based MPC. 
Note that in automotive applications, a feedback tracking controller is often coupled with a feedforward controller to speed up the transient response (\cite{norouzi2021model}).~ 
One of the most common feedforward controllers is based on look-up tables for set-points and actuator positions, however, an MPC-based feedforward~(\cite{liao2020model,ZHANG2022181}) has been shown to improve transient response.  

To support controller implementation, we first develop a multi-layered Neural Network (NN) model for engine-out $NOx$ and $Soot$ emissions based on experimental dynamometer data. Then, we integrate this NN-based emissions model with a high-fidelity engine model in GT-Power that predicts flows, pressures, and temperatures in the engine to obtain a high-fidelity model for closed-loop simulations. We then use this high-fidelity model to identify a low-order linear parameter-varying (LPV) model for emissions prediction, and we use this LPV model as a prediction model to implement supervisory economic MPC. Closed-loop simulation results are finally reported followed by concluding remarks.  



\section{Diesel Engine Modeling} \label{sec:model}
\begin{figure}[ht!]
\centering
\includegraphics[width=0.4\textwidth]{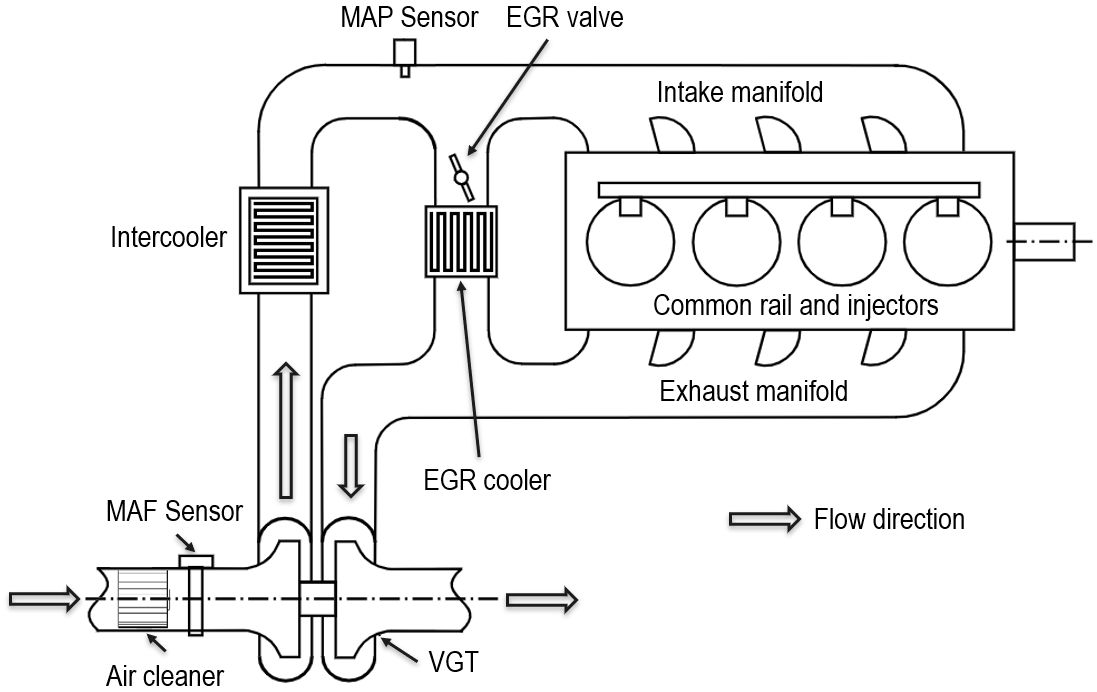}
\caption{Diesel engine with exhaust gas recirculation (EGR) and a variable geometry turbocharger (VGT).}
\label{fig:engine}
\end{figure}

\vspace{-3pt}
The schematic of the diesel engine considered in this paper is shown in Figure~\ref{fig:engine}. 
The EGR system dilutes the cylinder charge and lowers the peak combustion temperature, thereby reducing oxides of nitrogen $NOx$ emissions. The flow from the exhaust manifold of the engine to the intake manifold is controlled by the EGR valve. As the EGR flow also depends on the exhaust pressure, this flow is also affected by the VGT. Both VGT and EGR valve affect the intake manifold pressure and air-to-fuel ratio. High levels of recirculated exhaust gas and low air-to-fuel engine can lead to increased smoke/soot emissions. The engine operating point is typically defined by the engine speed ($N_{\tt e}$) and total fuel injection rate ($w_{\tt inj}$) inferred from the engine torque request; these are treated as the 
 exogenous inputs.
 
A typical diesel engine airpath controller tracks target values (set-points) for the intake manifold pressure ($p_{\tt im}$) and EGR rate ($\chi_{\tt egr}$). The EGR rate is defined by
\begin{equation} \label{eq:EGR_rate_def}
\chi_{\tt egr} = \frac{w_{\tt egr}}{w_{\tt egr} + w_{\tt c}},
\end{equation}
where $w_{\tt c}$ is the mass flow into the intake manifold through the compressor and intercooler,
and $w_{\tt egr}$ is the mass flow through the EGR valve from the exhaust manifold into the intake manifold. 

In this paper, the objective is to coordinate the EGR valve and VGT actuators to control the intake manifold pressure ($p_{\tt im}$) and EGR rate ($\chi_{\tt egr}$) to target values that are computed and provided to the airpath controller by a supervisory controller to satisfy emissions and fuel consumption requirements.  

As in our previous work on diesel engine airpath control (\cite{9867167,ZHANG2022181}), our controller development relies on a high-fidelity diesel engine model in GT-Power. This model represents the responses of flows, and pressures in different parts of the engine at a crankangle resolution and has been validated against experimental data from engine dynamometer testing. This model, however, does not represent engine feedgas emissions; hence, to enable emissions-oriented controller development in this paper, we first augment this GT-Power engine model with data-driven emissions models. We then develop control-oriented models for feedgas emissions and airpath, and use them for EMPC design.


%
\subsection{Data-driven Engine Feedgas Emissions Modeling}\vspace{-3pt}
%

\begin{figure}[ht!]
\centering
\includegraphics[width=0.4\textwidth]{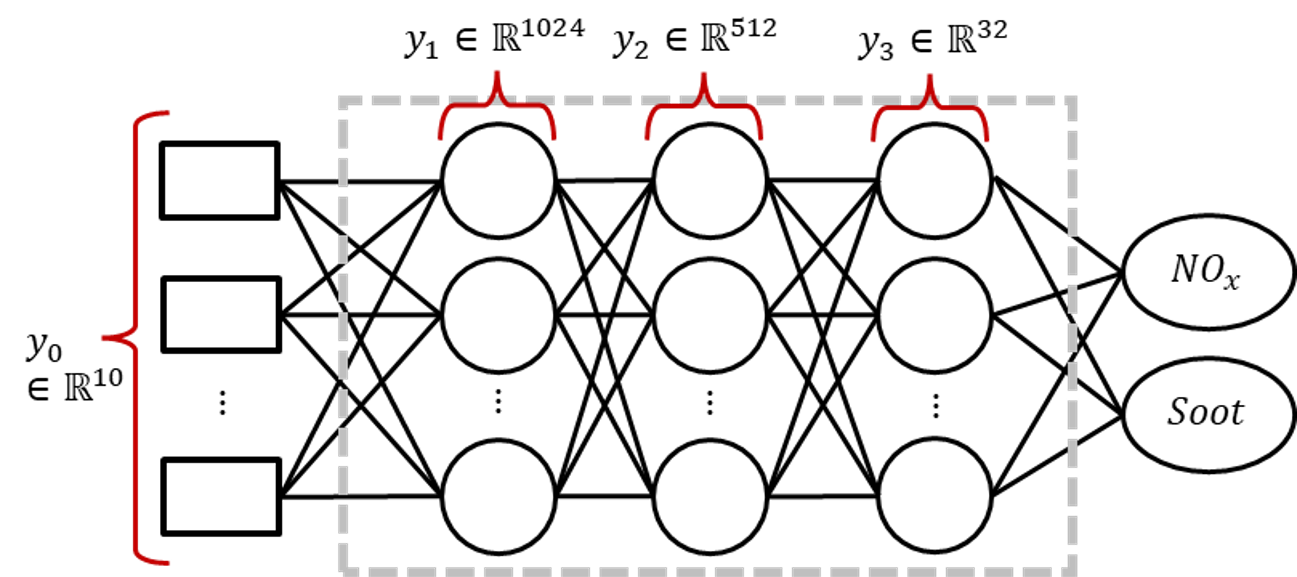}
\caption{Schematic of the developed multi-layer neural network diesel emissions model.}
\label{fig:nn_model}
\end{figure}

Neural Networks, as universal function approximators, are capable of learning emission models from data. Given experimental dynamometer data sets with both transient and steady-state measurements, we model the diesel engine emissions using a multi-layered NN (Figure~\ref{fig:nn_model}), 
where the output of each layer is defined by
\begin{equation}\label{eq:nn_plant}
    y_i=\sigma_{ReLU}\left(W_iy_{i-1}+b_{i}\right),~~ i=1,\dots,4,
\end{equation}
and where $W_i$ and $b_i$ are the network parameters of the $i${th} layer while $y_{0} \in\mathbb{R}^{10}$ and $y_{4}\in\mathbb{R}^{2}$ are the inputs and predicted emissions outputs, respectively.~ 
The NN is trained on an experimental dataset consisting of both steady-state ($\tt ss$) $\{y_{0,i}^{\tt ss}, y_{4,i}^{\tt ss}\}_{i=1,\dots,306}$ and transient ($\tt ts$) measurements $\{y_{0,i}^{\tt ts}, y_{4,i}^{\tt ts}\}_{i=1,\dots,12001}$.  The objective of the training a NN model is to minimize the sums of squares of the errors between the output $y_{4,i}$, $i=1,2$, and the actual emissions, namely, oxides of nitrogen ($NOx$) and $Soot$.   

The three hidden layers of
our NN model of size $1024$, $512$, and $32$, respectively, are chosen consistently with an encoder-decoder architecture while keeping the number of trainable parameters relatively small. The first hidden layer, functioning as an encoder, blends the input measurements from $\mathbb{R}^{10}$ into features in a higher dimension $\mathbb{R}^{1024}$. The consecutive two hidden layers have $512$ and $32$ neurons, respectively. They decode the high-dimensional features and project the vector layer-by-layer into $\mathbb{R}^{2}$.

\begin{table}[!htp]\centering
\caption{Normalized cross-covariance with zero lag between measured input variable candidates and emissions.}\label{tab:xcov}
\scriptsize
\begin{tabular}{l|cc|cc}\hline\toprule
 &\multicolumn{2}{c}{Transient} &\multicolumn{2}{c}{Steady State} \\\cmidrule{2-5}
Measured & $NOx$ &$Soot$ & $NOx$ &$Soot$ \\
Variables & $\rm [ppm]$ &$\rm [\%]$ & $\rm [ppm]$ &$\rm [\%]$ \\\cmidrule{1-5}
Injection Pressure $\rm [MPa]$&0.45 &-0.32 &-0.09 &-0.18 \\\cmidrule{1-5}
Main injection &\multirow{2}{*}{0.27} &\multirow{2}{*}{-0.39} &\multirow{2}{*}{-0.03} & \multirow{2}{*}{-0.19 } \\
timing $\rm [BTDC]$ &  & & &  \\\cmidrule{1-5}
Main injection fuel &\multirow{2}{*}{0.68} &\multirow{2}{*}{-0.12} &\multirow{2}{*}{0.38} & \multirow{2}{*}{-3.0e-3} \\
flow rate $\rm [mm^3/st]$ &  & & &  \\\cmidrule{1-5}
Pre-injection fuel &\multirow{2}{*}{0.22} &\multirow{2}{*}{0.07} &\multirow{2}{*}{-8.2e-17} & \multirow{2}{*}{-1.7e-16} \\
flow rate $\rm [mm^3/st]$ &  & & &  \\\cmidrule{1-5}
Engine torque output $\rm [Nm]$&0.67 &-0.10 &0.38 &-0.02 \\\cmidrule{1-5}
Engine speed $\rm [rpm]$&0.28 &-0.40 &-0.25 &-0.07 \\\cmidrule{1-5}
Intake manifold &\multirow{2}{*}{0.47} &\multirow{2}{*}{-0.21} &\multirow{2}{*}{0.15} & \multirow{2}{*}{0.15} \\
pressure $\rm [kPa]$ &  & & &  \\\cmidrule{1-5}
Exhaust manifold &\multirow{2}{*}{0.49} &\multirow{2}{*}{-0.22} &\multirow{2}{*}{0.11} & \multirow{2}{*}{-0.04} \\
pressure $\rm [kPa]$ &  & & &  \\\cmidrule{1-5}
Mass air flow $\rm [G/s]$&0.39 &-0.27 &0.04 &-0.12 \\\cmidrule{1-5}
EGR position $\rm [\%]$&-0.23 &-0.33 &0.10 &0.22 \\\cmidrule{1-5}
VGT position $\rm [\%]$&-0.32 &0.42 &0.08 &0.13 \\
\bottomrule\hline
\end{tabular}
\end{table}

To enable our NN model  to exploit two datasets (i.e., steady-state dataset and transient dataset) of distinct physical nature and dataset sizes we used the following procedure.
Firstly, the input variables used in training were selected based on correlation analysis. 
For instance,
Table~\ref{tab:xcov} indicates low normalized cross-covariance for the 
pre-injection fuel flow rate; hence we remove it from the set of model inputs 
while other signals in Table~\ref{tab:xcov} are kept in $y_0 \in\mathbb{R}^{10}$. 
Secondly, we exclude outliers from the transient training data set to facilitate the integration of the transient and  steady state data. For this, we empirically choose a threshold, $\epsilon>0$,  and exclude a transient data point $(y_{0,i}^{\tt ts}, y_{4,i}^{\tt ts})$ if  \vspace{-4pt}
\begin{equation}\label{eq:mahalanobis_r}
    r(y_{0,i}^{\tt ts}) = \left((y_{0,i}^{\tt ts}-\mu_{ss})^T\Sigma_{ss}^{-1}(y_{0,i}^{\tt ts}-\mu_{ss})\right)^{1/2} > \epsilon,
\end{equation}
where $r(y_{0,i}^{\tt ts})$ designates
the Mahalanobis distance, while
$\mu_{ss}$ and $\Sigma_{ss}$ stand for the mean and the covariance matrix of $\{y_{0,i}^{\tt ss}\}_{i=1,\dots,306}$. 
Thirdly, note that the transient dataset of $12001$ data points is significantly larger than the steady-state dataset of $306$ data points. To avoid overemphasizing transient response in training the NN model, the final training dataset comprises the transient dataset after outlier removal and the steady state dataset being duplicated sevenfold. 


Using the {\tt Pytorch} package developed by~\cite{pytorch}, the NN model was trained for $1000$ epochs with MSE-loss function and a batch size of $40$. The stochastic gradient descent algorithm with momentum $\rho=0.9$ was used for training. The initial learning rate is set to  $lr=10^{-4}$ with a decay rate of $\gamma=0.5$ for every $100$ epochs. The number of data points used for training, validation, and testing were chosen in the ratio of 70\%:15\%:15\%.

\begin{figure}[t]
\centering
\includegraphics[width=0.42\textwidth]{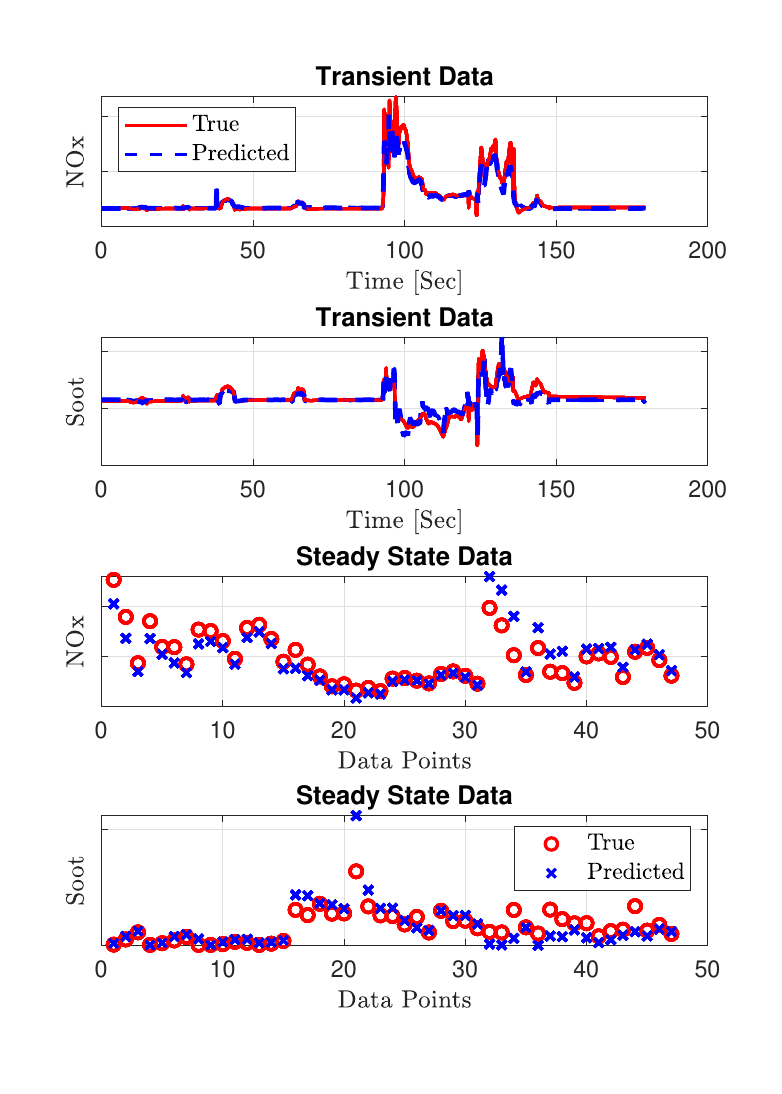}\vspace{-6pt}
\caption{The NN-based diesel emissions prediction performance against the actual dynamometer data.}
\label{fig:nn_prediction}
\end{figure}

Figure~\ref{fig:nn_prediction} compares the prediction results using the NN model versus the actual measured emissions data from the testing fraction of the dynamometer dataset. In Figure~\ref{fig:nn_prediction} and subsequent figures, we are not able to report the $y$-axis values in order to protect OEM proprietary data. Compared to the measured $NOx$ data, the predicted $NOx$ has mean errors of 49.8 $\rm ppm$ during transients and 129.8 $\rm ppm$ in steady state. The mean errors in $Soot$ prediction are 0.45\% in transients and 5.7\% in steady state. As shown in Figure~\ref{fig:nn_prediction}, our predictions for both transient data (in blue dash lines) and steady-state data (in blue crosses) match those from the dynamometer data with relatively small errors on average.
%

\vspace{-6pt}
\subsection{Emissions Control-oriented Modeling}\label{subsec:emission control-oriented model}\vspace{-3pt}
%
To reduce diesel engine emissions, we employ an economic MPC (EMPC). To implement EMPC,  a control-oriented model is needed for predicting the emissions over the MPC optimization horizon.~
To that end, in this section, we develop and validate an LPV model for emissions. To keep the prediction model as simple as possible, only $NOx$ and $Soot$ are selected as the LPV model states ($x$). Both emission outputs are assumed to be measured or accurately estimated.~ 
The input ($u$) for our model includes the intake manifold pressure ($p_{\tt im}$), EGR rate ($\chi_{\tt egr}$) and fuel injection rate ($w_{\tt inj}$).

Both the states and the input variables have been normalized by the steady-state values ($x^{\tt ss}(\rho)$, $u^{\tt ss}(\rho)$) and their corresponding standard deviation ($\sigma_{x}(\rho)$, $\sigma_u(\rho)$) as follows
\begin{equation} \label{eq: normalization}
\tilde{x}(\rho) = \frac{x-x^{\tt ss}(\rho)}{\sigma_{x}(\rho)},\quad \tilde{u}(\rho)=\frac{u-u^{\tt ss}(\rho)}{\sigma_{u}(\rho)}.
\end{equation}
The prediction model has an LPV form,
\begin{equation}  \label{eq:LPV model}
\tilde{x}_{k+1}(\rho_k) = \tilde{A}(\rho_k)\tilde{x}_k(\rho_k)
+\tilde{B}(\rho_k)\tilde{u}_k(\rho_k),
\end{equation}
where $k$ denote the discrete time, $\rho_k$ is the vector of engine speed and fuel injection rate at time instant $k$, $\tilde{A}: \mathbb{R}^2 \to \mathbb{R}^{2\times2}$, $\tilde{B}: \mathbb{R}^2 \to \mathbb{R}^{2\times3}$ are operating condition ($\rho$) dependent
matrices. 

\begin{figure}[ht!]
\centering
\includegraphics[width=1\columnwidth]{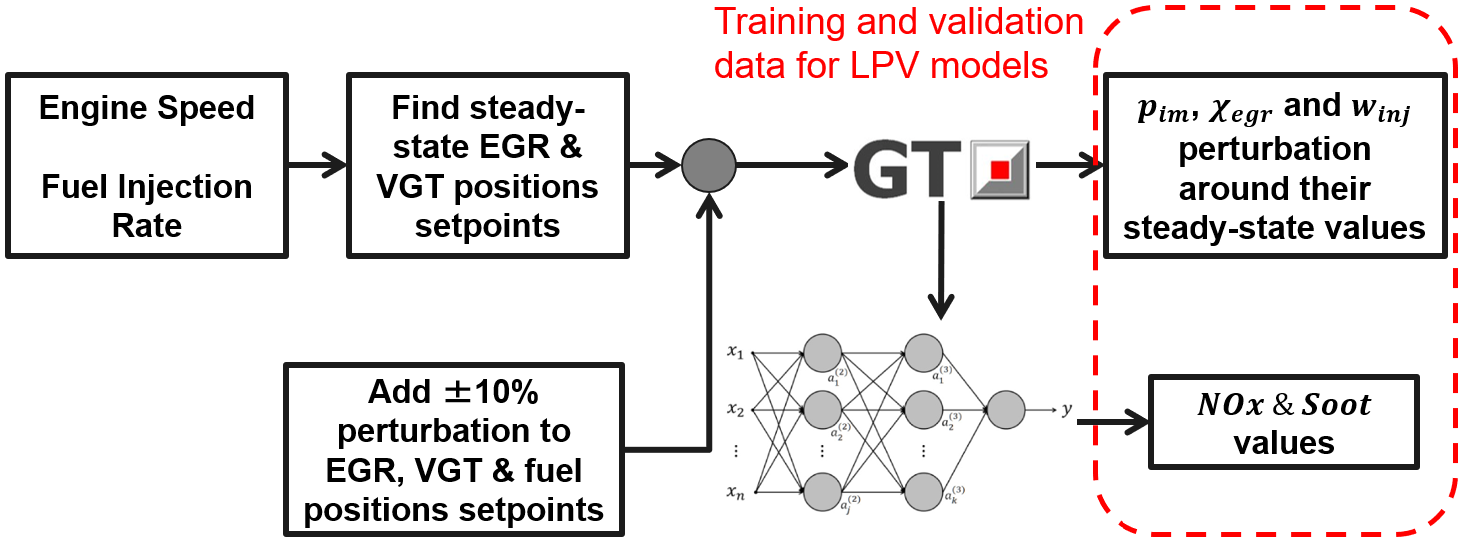}
\caption{Data generation procedure for identification of control-oriented LPV model of diesel emissions, where $p_{\tt im}$ is intake manifold pressure, $\chi_{\tt egr}$ is EGR rate, and $w_{\tt inj}$ is fuel injection quantity.}
\label{fig:LPV SID}
\end{figure}

We linearly interpolate the matrices $\tilde{A}(\rho)$ and $\tilde{B}(\rho)$ from $99$ models identified at pre-selected operating points ($\rho$) defined by 9 values of the engine speed and 11 values of fuel injection rate that cover the engine operating range. A co-simulation framework is implemented to integrate the NN emissions model with a model of the diesel engine airpath system built in GT-Power in order to generate input-output response data corresponding to small EGR and VGT position perturbations; these data are then used for local model identification at each of the $99$ operating points. The data generation procedure is summarized in Figure~\ref{fig:LPV SID}. {Using the function ${\tt n4sid}$ in MATLAB with prediction horizon setting as 50 steps, we could get the optimized $\tilde{A}$ and $\tilde{B}$ matrix for each operating point that minimizes the one-step-ahead prediction error between measured and predicted outputs.}

\begin{figure}[h!]
\centering
\includegraphics[width=0.48\textwidth]{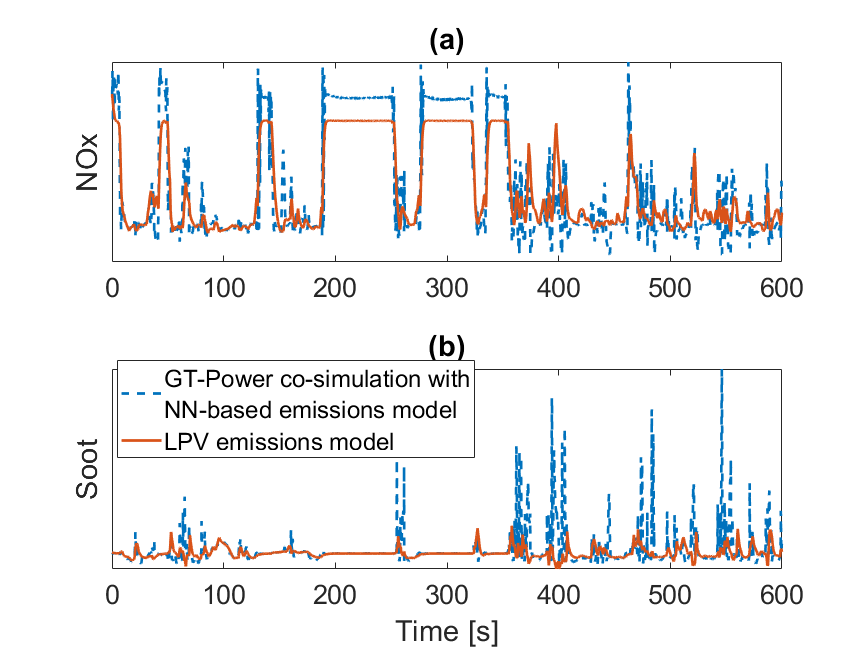}\vspace{-6pt}
\caption{The comparison results between control-oriented LPV emissions model against the GT-Power model co-simulation with NN-based emissions model over the first 600 $s$ of WHTC driving cycle: (a) $NOx$, and (b) $Soot$.}\vspace{-5pt}
\label{fig:LPV validation}
\end{figure}

Figure~\ref{fig:LPV validation} compares the results of simulating the LPV model \eqref{eq:LPV model} and the GT-Power model with NN-based emissions model over the first $600s$ of the World Harmonized Transient Cycle (WHTC) driving cycle, where both models respond to the same input trajectories of engine speed, fuel injection rate, target intake manifold pressure, and target EGR rate. The trajectories of the target intake manifold pressure and target EGR rate for these simulations were generated from
look-up tables that specify desired $p_{\tt im}$
and $\chi_{\tt egr}$ in steady-state as functions of the engine speed and fuel injection rate. 

The average errors in $NOx$ and $Soot$ LPV models as compared to the higher-fidelity NN model are less than $100~ppm$ and 2\%, respectively, even though larger errors are observed in transients. While the transient errors could potentially be reduced with higher-order LPV or/and NN models, as we will demonstrate in the following section, the proposed simple LPV model could successfully support an EMPC design.
\vspace{-4pt}

\subsection{Diesel Airpath Control-Oriented Model}\label{subsec:airpath control-oriented model}
A control-oriented model of the diesel airpath system is adopted to design the inner-loop tracking MPC. The airpath model estimates the response of the intake manifold pressure and EGR rate to EGR and VGT actuators. The intake manifold pressure ($p_{\tt im}$) and EGR rate ($\chi_{\tt egr}$) are two of the airpath model states ($z$). The model inputs ($v$) are the EGR valve position (percent open) and VGT position (percent close). The control-oriented airpath model also has an LPV form,
\begin{gather} \label{eq:airpath LPV model}
z_{k+1} - z_{k+1}^{\tt ss}(\rho_k)  =  A(\rho_k) \left[z_k-z_k^{\tt ss}(\rho_k)\right] \nonumber \\  +  
B(\rho_k) \left[v_k-v_k^{\tt ss}(\rho_k)\right] 
+B_f(\rho_k) \left[w_{{\tt inj},k}-w_{{\tt inj}}^{\tt ss}(\rho_k)\right],
\end{gather}
where~
$A,B: \mathbb{R}^2 \to \mathbb{R}^{2\times2}$,
$z_k^{\tt ss}, v_k^{\tt ss}: \mathbb{R}^2 \to \mathbb{R}^2$ are mappings that determine equilibrium values of $z$ and $v$ corresponding to a given $\rho$,  $B_f:~ \mathbb{R}^2 \to \mathbb{R}^{2 \times 1}$, and  $w_{\tt inj}^{\tt ss}(\rho)=\left[\begin{array}{cc} 0 & 1 \end{array} \right] \rho$. Note that the model (\ref{eq:airpath LPV model}) includes $w_{{\tt inj},k}$ as an extra additive input. This input has been added as includes it improves the model match in transients. Details of the airpath LPV model identification and validation can be found in \cite{ZHANG2022181}.

\section{Integrated Emissions and Airpath MPC Design}\label{sec:controller design}\vspace{-3pt}

\begin{figure}[ht!]
\centering
\includegraphics[width=0.45\textwidth]{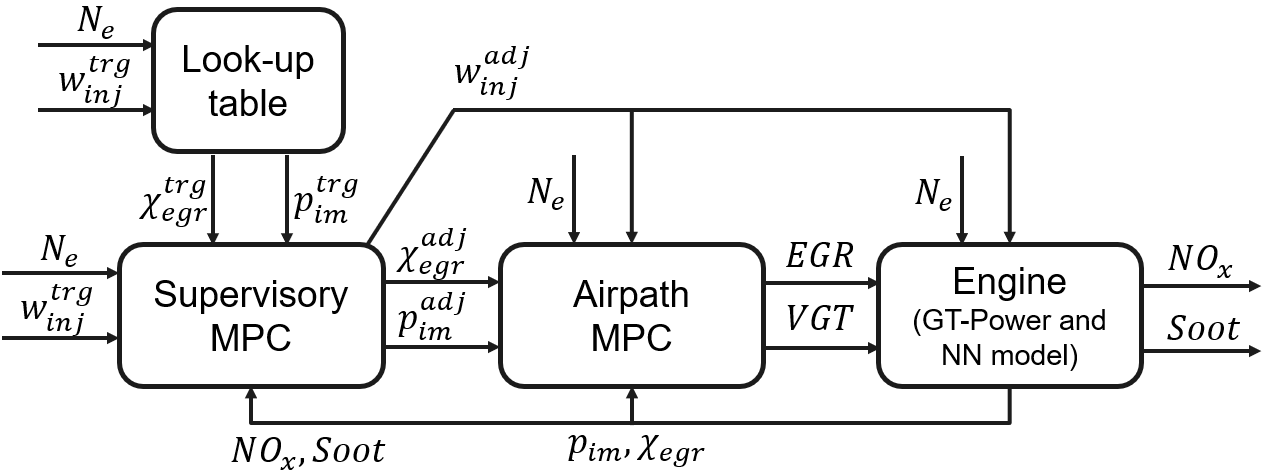}
\caption{The block diagram of the integrated emissions and airpath control system. 
}
\label{fig:structure}
\end{figure}

The overall architecture of the proposed integrated emissions and airpath controller is shown in Figure~\ref{fig:structure}.~
For each engine speed ($N_{\tt e}$) and fuel injection rate target ($w_{\tt inj}$), the outer-loop EMPC controller generates adjusted target values for intake manifold pressure ($p_{\tt im}^{adj}$), EGR rate ($\chi_{\tt egr}^{adj}$), as well as the fuel injection command ($w_{\tt inj}^{adj}$) to reduce $NOx$ and enforce the constraint on $Soot$. The fuel injection rate is applied directly to the engine while the adjusted intake manifold pressure and EGR rate target values are passed to the airpath
MPC which tracks these targets by adjusting the EGR valve and VGT positions, see  \cite{ZHANG2022181} for more details.

In the baseline engine control strategy, the intake manifold pressure and EGR rate targets are static functions of the operating condition, i.e., $p_{\tt im}^{\tt trg}(\rho): \mathbb{R}^2 \to \mathbb{R}$ and $\chi_{\tt egr}^{\tt trg}(\rho): \mathbb{R}^2 \to \mathbb{R}$ that are implemented using look-up tables. These targets are obtained during engine development and are chosen to be optimal in steady-state operating conditions. In this paper, we focus on the optimization of transient engine response, i.e., using MPC to shape the transient response of the engine as it transitions between operating points.

\vspace{-4pt}
\subsection{EMPC for Emissions Control}
{
Our economic MPC (EMPC) exploits a rate-based prediction model to enhance closed-loop robustness and disturbance rejection.} 
Rate-based MPC was applied to aircraft turbofan engines by \cite{decastro2007rate} and to diesel engine airpath control by \cite{huang2013rate,huang2014robust,huang2016rate}.  

First, in reference to the model (\ref{eq:LPV model}), we define rate 
variables $\Delta x_{k}$ and $\Delta u_{k}$ as $$\Delta x_{k} = \tilde{x}_{k} - \tilde{x}_{k-1}, \quad \Delta u_{k} = \tilde{u}_{k} - \tilde{u}_{k-1},$$
that correspond to the increments in the state and control variables, respectively. Then, assuming $\rho_k$ remains constant over the prediction horizon, the prediction model (\ref{eq:LPV model}) implies that
\begin{equation} \label{eq:rate base emission model}
\Delta x_{k+1} = \tilde{A}(\rho_k) \Delta x_k + \tilde{B}(\rho_k) \Delta u_k.
\end{equation}
The constant $\rho_k$ assumption is adopted widely in the literature on engine and powertrain control with MPC (\cite{norouzi2021model}).

The rate-based reformulation as given in \eqref{eq:rate base emission model} is particularly effective when applied to an LPV model, as the steady-state values of states and controls in \eqref{eq: normalization}, i.e., $x_k^{ss}(\rho_k)$
and $u_k^{ss}(\rho_k)$, under the assumption of $\rho_k$ remaining constant over the prediction horizon, does not need to be known.

We let $\tilde{x}_{j|k}$ and $\tilde{u}_{j|k}$ denote the predicted state and control values at time step $j,$
$0 \leq j \leq N$, over the prediction horizon when the prediction is made at the time step $k$. Then
to be able to impose constraints on $\tilde{x}_{j|k}$ and $\tilde{u}_{j|k}$ and compute the MPC cost that penalizes the rate of change of input, $\Delta u_{k}$, we define
the augmented state vector,
$$x_{j|k}^{\tt ext} = \left[\begin{array}{cccc} \Delta x_{j|k}^{\sf T}, & \tilde{x}_{j-1|k}^{\sf T}, & \tilde{u}_{j-1|k}^{\sf T} \end{array}\right]^{\sf T}.$$
The rate-based prediction model \eqref{eq:rate base emission model} then implies
\begin{align}
    &x_{j+1|k}^{\tt ext} = \begin{bmatrix}
    \tilde{A}(\rho_k) & 0 & 0\\
    \mathbb{I}_{n_x\times n_x} & \mathbb{I}_{n_x\times n_x} & 0\\
    0 & 0 & \mathbb{I}_{n_u\times n_u}
    \end{bmatrix}
    x_{j|k}^{\tt ext}  \nonumber\\ 
    & \qquad \qquad + \begin{bmatrix}
    \tilde{B}(\rho_k) \\
    0\\
    \mathbb{I}_{n_u\times n_u}
    \end{bmatrix} \Delta u_{j|k},
\end{align}
where $\Delta u_{j|k}$ is the control input in the extended system.

The EMPC is designed based on considerations of safety and drivability, so that
the fuel injection rate satisfies the specified upper and lower bounds, as well as on the considerations of the emission requirements such as $Soot$ limits.
The following optimal control problem is solved at each sampling instant with the intake manifold pressure, EGR rate, and fuel injection rate being optimized:
\begin{subequations}\label{eq:modified emission MPC}
\begin{multline}
    \min_{\Delta u_{0|k},...,\Delta u_{N-1|k},\epsilon_k} J(\Delta u,\epsilon,\rho) = \sum_{j=0}^{N} l(\Delta u_{j|k},\epsilon_{j|k},\rho_k)
\tag{\ref{eq:modified emission MPC}}
\end{multline}

subject to
\begin{align} 
    &\Delta x_{j+1|k} = \tilde{A}(\rho_k)\Delta x_{j|k} + \tilde{B}(\rho_k)\Delta u_{j|k}\\
    &u_{j|k} = u_{j-1|k} + \Delta u_{j|k}, j = {0}, \ldots, N-1,\\ 
    &{Soot}_{j|k} \leq {Soot}_{max} + \epsilon_{j|k}, j = 1, \ldots, N,\\
    &0.9w_{\tt inj}^{trg}\leq {w_{\tt inj}^{adj}}_{j|k} \leq w_{\tt inj}^{trg}, j = 0,\ldots, N-1,\\
    &\epsilon_{j|k} \geq 0, j = 0,\ldots, N-1,
\end{align}
\end{subequations}

where $N$ is the prediction horizon and $\epsilon_k$ is the slack variable introduced to avoid infeasibility of the $Soot$ constraints. The stage cost function $l$ is defined as
\begin{align*}
    l(\Delta u,\epsilon,\rho) = &\alpha(p_{\tt im}^{trg}(\rho)-p_{\tt im}^{adj})^2 + \beta(\chi_{\tt egr}^{trg}(\rho)-\chi_{\tt egr}^{adj})^2 + \\
    &\gamma(w_{\tt inj}^{trg}-w_{\tt inj}^{adj}) + \eta {NOx} + \zeta \epsilon + \Delta u^{\sf T}R\Delta u
\end{align*}
where $\alpha,\beta,\gamma,\eta,\zeta > 0$ and $R > 0$ are tuning parameters and reflect tracking objectives for the intake manifold pressure target, EGR rate target, and fuel injection rate, a penalty on $NOx$ value, a penalty to soften the $Soot$ constraint to guarantee feasibility, and a damping term. The fuel tracking, $NOx$ term, and slack variable use 1-norm penalties since they are more robust to ill-conditioning compared to quadratic penalties, facilitating the solution of \eqref{eq:modified emission MPC} numerically. 

\vspace{-6pt}
\subsection{Tracking MPC for Airpath System}
Our airpath MPC includes a feedforward (FF) MPC, to provide fast transient response, and a feedback (FB) MPC
that provides integral action and ensures offset-free tracking.
The FB MPC uses a rate-based formulation to eliminate the steady-state error while the FF MPC only closes the loop around the model of the plant.  Details of design and tuning of the airpath MPC are presented in \cite{ZHANG2022181}.

\section{Simulation Results and Discussions} \label{sec:results}\vspace{-3pt}
Four different tuning scenarios are considered to investigate how the emissions can be altered according to different control requirements:
\begin{itemize}
  \item \textbf{EMPC-A}: Low $NOx$ penalty without $Soot$ limit
  \item \textbf{EMPC-B}: High $NOx$ penalty without $Soot$ limit
  \item \textbf{EMPC-C}: Low $NOx$ penalty with $Soot$ limit
  \item \textbf{EMPC-D}: High $NOx$ penalty with $Soot$ limit
\end{itemize}
A baseline scenario without EMPC (\textbf{w/o EMPC}) was also generated, in which the adjusted targets for 
$p_{\tt im}$, $\chi_{\tt egr}$ are the same as the values from look-up table while $w_{\tt inj}$ is the same as defined by the operating condition.

\subsection{Case Study with Steps/Ramps}
\begin{figure}[h!]
\centering
\includegraphics[width=0.4\textwidth]{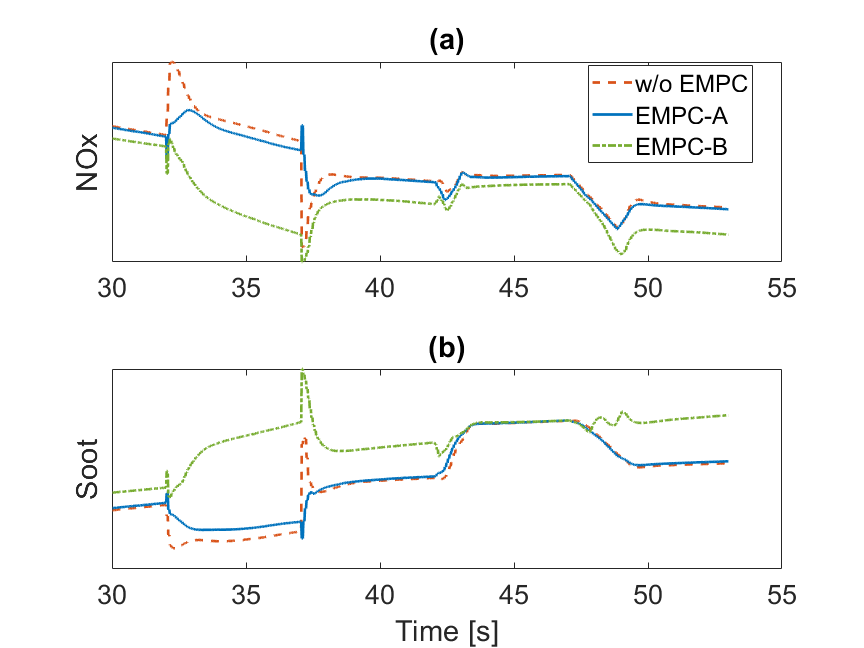}\vspace{-6pt}
\caption{Comparison of (a) $NOx$ and (b) $Soot$ with different $NOx$ penalties during {first, a fuel ($w_{\tt inj}$) tip-in and tip-out and then, a ramp change of $N_{\tt e}$.} 
}\vspace{-8pt}
\label{fig:case 1 NOx compare}
\end{figure}

\begin{figure}[h!]
\centering
\includegraphics[width=0.4\textwidth]{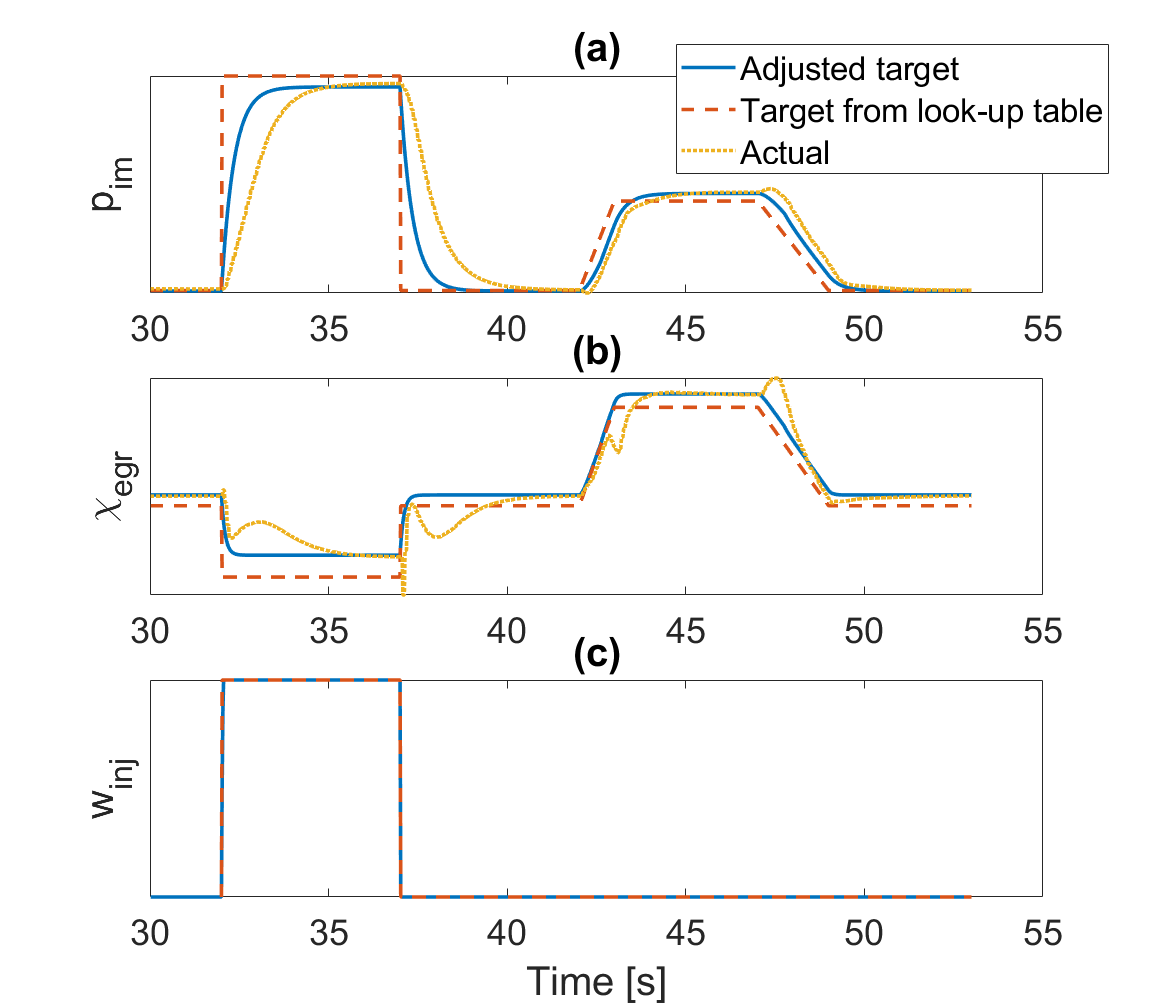}\vspace{-6pt}
\caption{Comparison of actual
, target from look-up table, and adjusted target by EMPC (a) intake manifold pressure, (b) EGR rate and (c) fuel injection rate with \textbf{EMPC-B}.}
\label{fig:case 1 input wo soot lim}
\end{figure}

\vspace{-3pt}
Figure~\ref{fig:case 1 NOx compare} shows the simulation results at one operating point with different $NOx$ penalties in the cost function during first, a fuel tip-in and tip-out and then, a ramp change of engine speed. Here no $Soot$ limit was added. {The simulations are run for 30 $s$ first to achieve close to steady-state temperature conditions before steps are applied.}  According to the results summarized in Table~\ref{tbl:case study NOx}, the EMPC  reduces both cumulative $NOx$ and peak $NOx$ values. Figure~\ref{fig:case 1 input wo soot lim} shows the corresponding intake manifold pressure, EGR rate, and fuel injection rate values from EMPC-B simulation. The EMPC adjusts the target $p_{\tt im}$ and $\chi_{\tt egr}$ to help reduce $NOx$ while the airpath MPC is able to track these adjusted target values without steady-state errors. 
\vspace{-6pt}
\vspace{-6pt}

\begin{figure}[h!]
\centering
\includegraphics[width=0.4\textwidth]{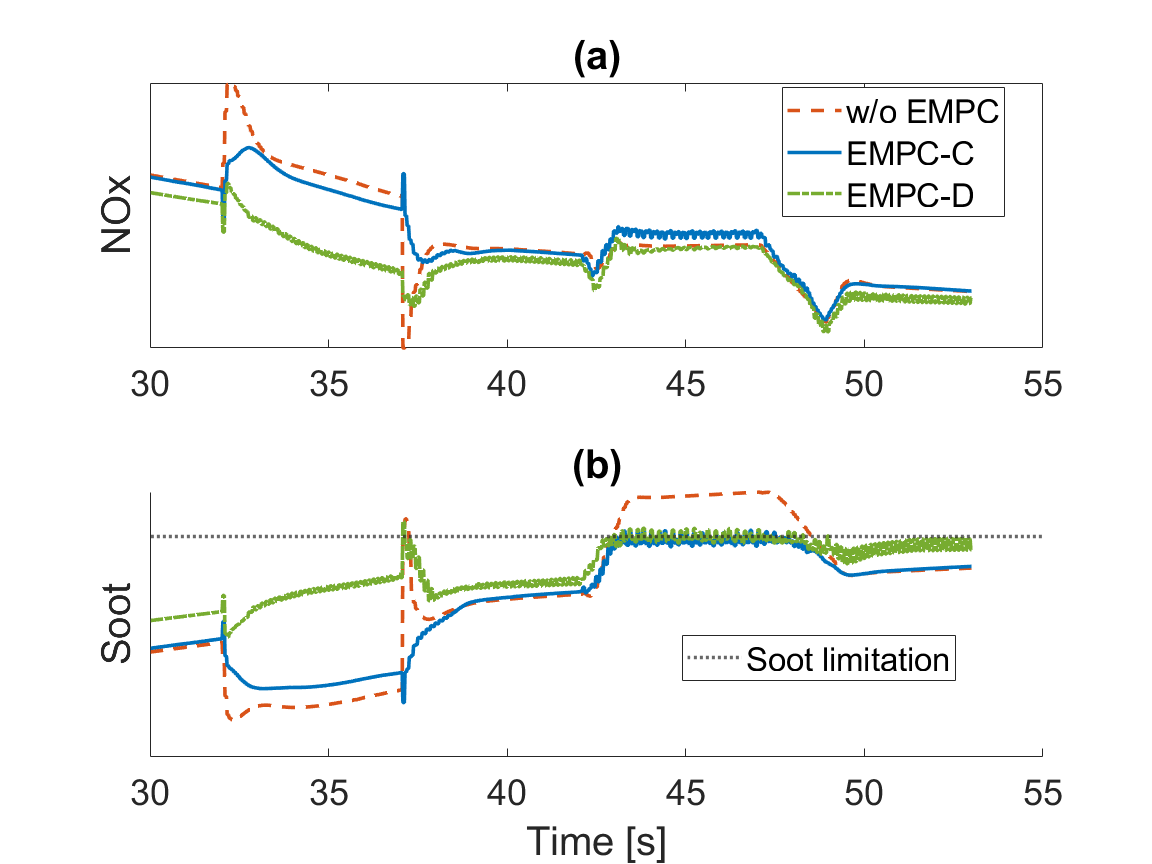}\vspace{-6pt}
\caption{Comparison of (a) $NOx$ and (b) $Soot$ with the same $NOx$ penalty and operating conditions as in Figure~\ref{fig:case 1 NOx compare} with the additional of $Soot$ limit.}\vspace{-7pt}
\label{fig:case 1 Soot compare}
\end{figure}

\begin{figure}[h!]
\centering
\includegraphics[width=0.4\textwidth]{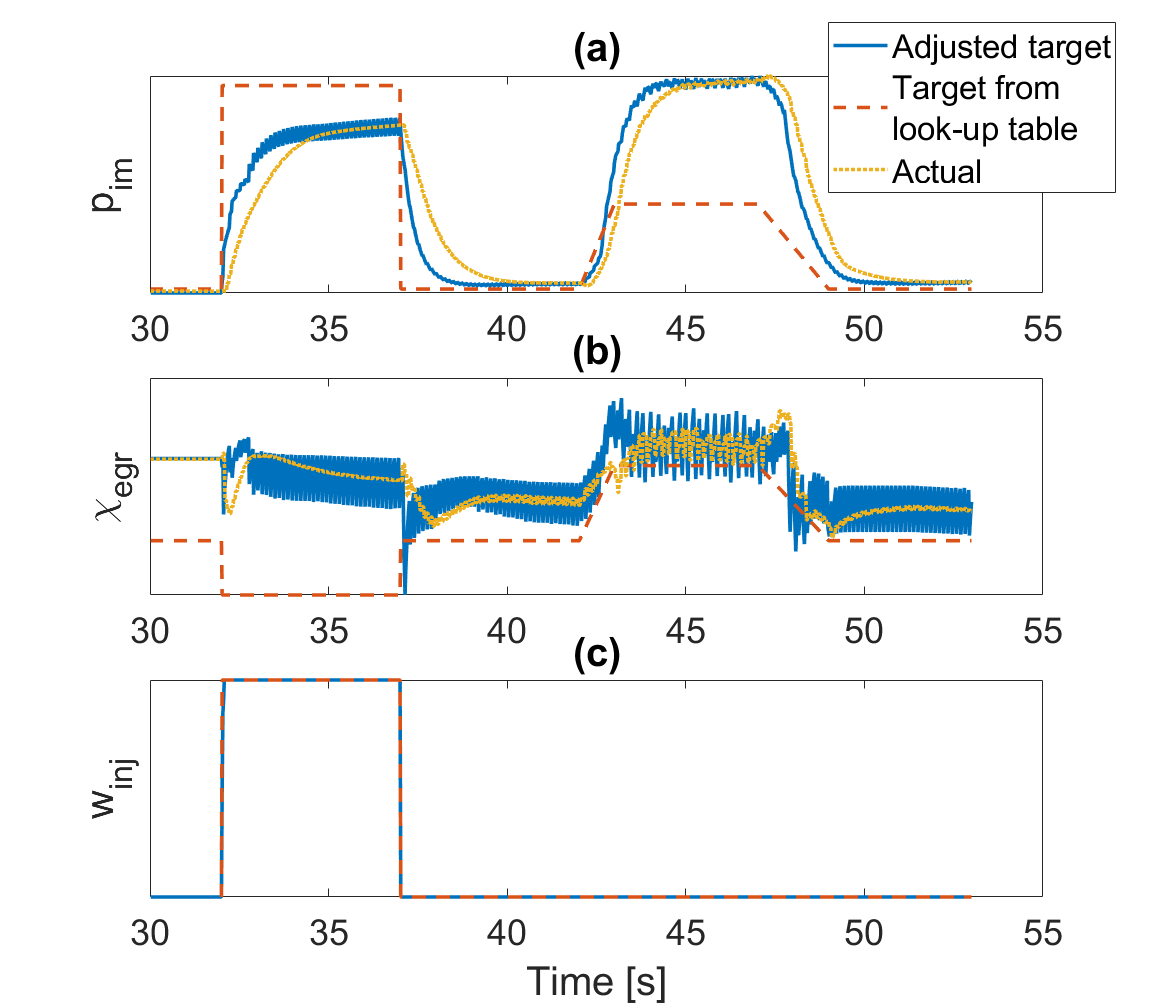}\vspace{-6pt}
\caption{Comparison of actual, target from look-up table, and adjusted target by EMPC (a) intake manifold pressure, (b) EGR rate, and (c) fuel injection rate with \textbf{EMPC-D}.}\vspace{-5pt}
\label{fig:case 1 input wi soot lim}
\end{figure}

After adding the $Soot$ constraint, the EMPC is able to keep $Soot$ below a  predefined limit. Figure~\ref{fig:case 1 Soot compare} shows the simulation results at the same operating point and with the $Soot$ limit, and with different $NOx$ penalties in the cost function.  Table~\ref{tbl:case study Soot} indicates that both EMPC-C and EMPC-D have reduced $Soot$ peak values. However, EMPC-D has  higher average $Soot$ compared with the baseline due to its large penalty on $NOx$ in the cost function.  According to Figure~\ref{fig:case 1 input wi soot lim}, $Soot$ values of both EMPCs are below the limit for most of the  time. Oscillations in the adjusted target values of $p_{\tt im}$ and $\chi_{\tt 
egr}$ are observed in Figure~\ref{fig:case 1 input wi soot lim} attributed to the model mismatch between LPV and NN emissions models. {The range of the oscillation in $p_{\tt im}$ and $\chi_{\tt 
egr}$ is less than 0.01 bar and 1\%, respectively, and is relatively small. 
Such oscillation can also be reduced by using a larger penalty $R$ on the damping term in the MPC cost function.}
At the same time, the airpath MPC successfully tracks the adjusted target values.
%
\vspace{-10pt}

\begin{table}[ht!]
\caption{Comparison of the $NOx$ results for different EMPC scenarios.}\vspace{-3pt}
\label{tbl:case study NOx}
{\scriptsize
\begin{center}
\begin{tabular}{lll}
\toprule  
\makecell[l]{\textbf{MPC}} & \makecell[l]{\textbf{Cumulative $NOx$ [\%]}} & \makecell[l]{\textbf{Peak $NOx$ [\%]}}\\
\midrule  
\textbf{w/o EMPC} & reference & reference\\ 
(baseline) &  & \\ \hline
\textbf{EMPC-A} & $\downarrow$ 0.780\% & $\downarrow$ 5.700\%\\\hline
\textbf{EMPC-B} & $\downarrow$ 6.695\% & $\downarrow$ 9.065\%\\\hline
 \textbf{EMPC-C} & $\uparrow$ 0.395\% & $\downarrow$ 5.341\%\\\hline
\textbf{EMPC-D} & $\downarrow$ 3.033\% & $\downarrow$ 8.301\%\\
\bottomrule 
\end{tabular}
\end{center}}
\end{table}

\begin{table}[h!]
\caption{Comparison of the $Soot$ results for different EMPC scenarios.}\vspace{-3pt}
\label{tbl:case study Soot}
{\scriptsize
\begin{center}
\begin{tabular}{lll}
\toprule  
\makecell[l]{\textbf{MPC}} & \makecell[l]{\textbf{Average $Soot$ [\%]}} & \makecell[l]{\textbf{Peak $Soot$ [\%]}}\\
\midrule  
\textbf{w/o EMPC} & reference & reference\\ 
(baseline) &  & \\ \hline
\textbf{EMPC-A} & $\uparrow$ 1.067\% & $\uparrow$ 0.090\%\\\hline
\textbf{EMPC-B} & $\uparrow$ 14.071\% & $\uparrow$ 14.946\%\\\hline
 \textbf{EMPC-C} & $\downarrow$ 1.300\% & $\downarrow$ 6.183\%\\\hline
\textbf{EMPC-D} & $\uparrow$ 5.003\% & $\downarrow$ 4.765\%\\
\bottomrule 
\end{tabular}
\end{center}}
\end{table}
 
\subsection{Transient Drive Cycle Simulations}
\begin{figure}[h!]
\centering
\includegraphics[width=0.45\textwidth]{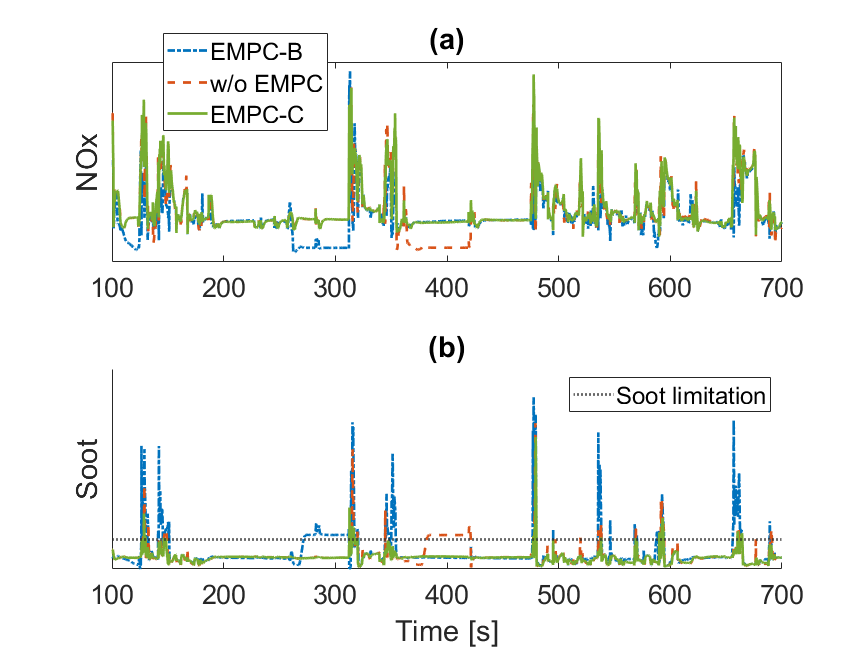}\vspace{-5pt}
\caption{Comparison of (a) $NOx$ and (b) $Soot$ control results over the FTP cycle.}\vspace{-10pt}
\label{fig:FTP compare}
\end{figure}

\begin{figure}[h!]
\centering
\includegraphics[width=0.45\textwidth]{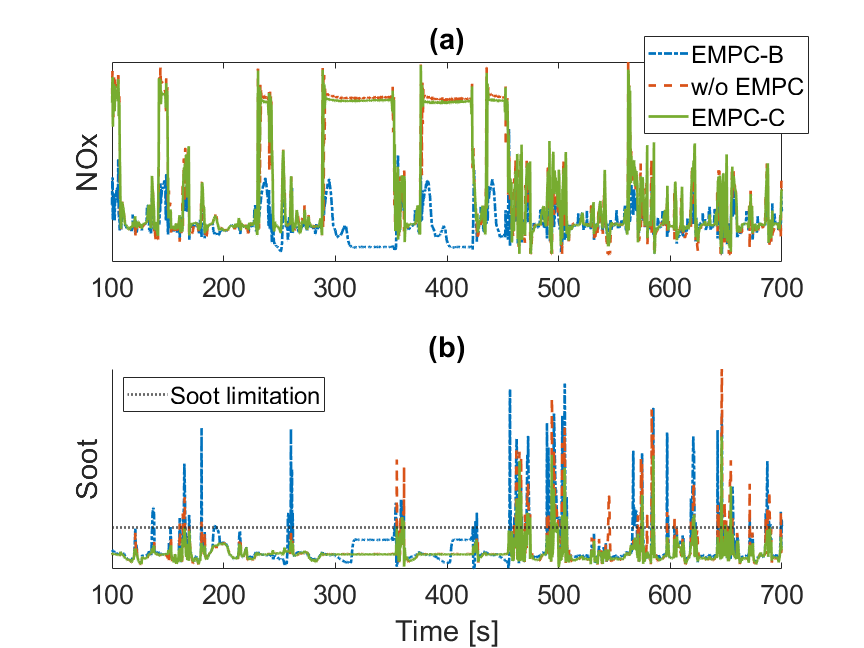}\vspace{-5pt}
\caption{Comparison of (a) $NOx$ and (b) $Soot$ control results over the WHTC cycle.}\vspace{-7pt}
\label{fig:WHTC compare}
\end{figure}

We now evaluate the integrated emissions and airpath control strategy in more complex simulation scenarios over the Federal Test Procedure (FTP) and 
WHTC. The results are summarized in Tables~\ref{tbl:cycle NOx} and \ref{tbl:cycle Soot}. {Both cycles run for 100 $s$ first to achieve close to steady-state temperature conditions before cycle inputs are applied.} Figures~\ref{fig:FTP compare} and \ref{fig:WHTC compare} show the comparison between EMPC-B and EMPC-C against the baseline case over FTP and WHTC cycles, respectively. With EMPC-B, $NOx$ are considerably reduced over both drive cycles with a sacrifice of around 20\% increase in average $Soot$. If the $Soot$ limit is imposed, EMPC-C could reduce both average and peak $Soot$ in both drive cycles compared with the baseline case. Moreover, according to Figures \ref{fig:FTP compare}-(b) and \ref{fig:WHTC compare}-(b), most of the peaks above the $Soot$ constraint are eliminated, illustrating EMPC-C is able to reduce the visible smoke. Results of EMPC-D are only shown in Tables~\ref{tbl:cycle NOx} and \ref{tbl:cycle Soot}, from which it can be seen that both cumulative $NOx$ and average $Soot$ are reduced if the EMPC is made very aggressive on both $NOx$ reduction and $Soot$ limits.\vspace{-6pt}  

\begin{table}[ht!]
\caption{Comparison of $NOx$ control results over FTP and WHTC cycles.}\vspace{-3pt}
\label{tbl:cycle NOx}
{\scriptsize
\begin{center}
\begin{tabular}{lll}
\toprule  
\makecell[l]{\textbf{MPC}} & \makecell[l]{\textbf{Cumulative $NOx$ [\%]}} & \makecell[l]{\textbf{Peak $NOx$ [\%]}}\\
\midrule  
\textbf{w/o EMPC} & reference & reference\\ 
(baseline) & FTP/WHTC & FTP/WHTC\\ \hline
 \textbf{EMPC-B} & $\downarrow$ 9.527\%/$\downarrow$ 38.644\% & $\uparrow$ 8.374\%/$\downarrow$ 33.530\%\\\hline
 \textbf{EMPC-C} & $\uparrow$ 4.056\%/$\uparrow$ 2.963\% & $\uparrow$ 6.6059\%/$\downarrow$ 1.649\%\\\hline
\textbf{EMPC-D} & $\downarrow$ 0.451\%/$\downarrow$ 15.378\% & $\uparrow$ 10.822\%/$\downarrow$ 11.615\%\\
\bottomrule 
\end{tabular}
\end{center}}
\end{table}

\begin{table}[ht!]
\caption{Comparison of $Soot$ control results over FTP and WHTC cycles.}\vspace{-3pt}
\label{tbl:cycle Soot}
{\scriptsize
\begin{center}
\begin{tabular}{lll}
\toprule  
\makecell[l]{\textbf{MPC}} & \makecell[l]{\textbf{Average $Soot$ [\%]}} & \makecell[l]{\textbf{Peak $Soot$ [\%]}}\\
\midrule  
\textbf{w/o EMPC} & reference & reference\\ 
(baseline) & FTP/WHTC & FTP/WHTC\\ \hline
 \textbf{EMPC-B}  & $\uparrow$ 18.789\%/$\uparrow$ 21.329\% & $\uparrow$ 17.800\%/$\downarrow$ 7.238\%\\\hline
 \textbf{EMPC-C} & $\downarrow$ 14.268\%/$\downarrow$ 9.521\% & $\downarrow$ 9.485\%/$\downarrow$ 33.786\%\\\hline
\textbf{EMPC-D}  & $\downarrow$ 3.262\%/$\downarrow$ 5.656\% & $\downarrow$ 9.147\%/$\downarrow$ 34.741\%\\
\bottomrule 
\end{tabular}
\end{center}}
\end{table}

\section{Conclusions}\label{sec:conclusion}

In this paper, a multi-layer Neural Network (NN) was trained and validated to simultaneously predict $NOx$ and $Soot$ emissions of a diesel engine for both transient and steady-state operating conditions. A co-simulation framework was then implemented to integrate the NN model with a high-fidelity model of the diesel engine airpath system built in GT-Power. An integrated control strategy for emissions regulation and airpath system setpoint tracking was developed based on model predictive control (MPC). The emission controller is based on an economic MPC formulation that leverages a linear parameter varying (LPV) model of $NOx$ and $Soot$ to minimize $NOx$ and enforce the $Soot$ limit. The airpath controller is a rate-based MPC consisting of feedback and feedforward loops, aiming to track adjusted intake manifold pressure and EGR rate setpoints computed by the economic MPC.


The performance of our integrated emissions and airpath MPCs was assessed through simulations over different steady-state and transient drive cycles in terms of (i) reducing $NOx$, (ii) enforcing $Soot$ limit, and (iii) tracking the adjusted intake manifold pressure and EGR rate setpoints.  The results highlight the potential of economic supervisory MPC to achieve improved transient control and for reducing engine feedgas emissions.~ 
Future works include using region-based EMPC calibration to improve transient performance and replacing the LPV model used for emission prediction with a dynamic neural network model to reduce model mismatch and improve the overall performance.

\bibliography{ifacconf}             

\end{document}